\documentclass[prd,aps,nofootinbib,twocolumn,superscriptaddress]{revtex4}

\usepackage{color,natbib,wrapfig,ulem}

\usepackage{graphics}
\usepackage{amsmath}
\usepackage{epsfig}
\usepackage{multirow}
\newcommand{\pbar}{$\bar{p}~$}

\newcommand{\dbar}{$\overline{\text{D}}~$}
\newcommand{\antihe}{$^3\overline{\text{He}}~$}
\newcommand{\antiT}{$^3\overline{\text{H}}~$}

\begin{document}

\title{\mbox{Antihelium from Dark Matter}}

\author{Eric Carlson}
\affiliation{Department of Physics, University of California, 1156 High St., Santa Cruz, CA 95064, USA}
\affiliation{Santa Cruz Institute for Particle Physics, Santa Cruz, CA 95064, USA} 
\email{erccarls@ucsc.edu}

\author{Adam Coogan}
\email{acoogan@ucsc.edu}\affiliation{Department of Physics, University of California, 1156 High St., Santa Cruz, CA 95064, USA}\affiliation{Santa Cruz Institute for Particle Physics, Santa Cruz, CA 95064, USA} 

\author{Tim Linden}
\email{trlinden@uchicago.edu}\affiliation{Department of Physics, University of California, 1156 High St., Santa Cruz, CA 95064, USA}\affiliation{Santa Cruz Institute for Particle Physics, Santa Cruz, CA 95064, USA} \affiliation{Department of Physics, University of Chicago, Chicago, IL 60637}\affiliation{Kavli Institute for Cosmological Physics, Chicago, IL 60637}

\author{Stefano Profumo}
\email{profumo@ucsc.edu}\affiliation{Department of Physics, University of California, 1156 High St., Santa Cruz, CA 95064, USA}\affiliation{Santa Cruz Institute for Particle Physics, Santa Cruz, CA 95064, USA} 

\author{Alejandro Ibarra}
\email{ibarra@tum.de}\affiliation{Physik-Department T30d, Technische Universit\"at M\"unchen, James-Franck-Stra\ss{}e, 85748 Garching, Germany}

\author{Sebastian Wild}
\email{sebastian-wild@mytum.de}\affiliation{Physik-Department T30d, Technische Universit\"at M\"unchen, James-Franck-Stra\ss{}e, 85748 Garching, Germany}

\date{\today}

\begin{abstract}
\noindent  Cosmic-ray anti-nuclei provide a promising discovery channel for the indirect detection of particle dark matter.  Hadron showers produced by the pair-annihilation or decay of Galactic dark matter generate anti-nucleons which can in turn form light anti-nuclei.  Previous studies have only focused on the spectrum and flux of low energy antideuterons which, although very rarely, are occasionally also produced by cosmic-ray spallation. Heavier elements ($A\geq3$) have instead entirely negligible astrophysical background and a primary yield from dark matter which could be detectable by future experiments.  Using a Monte Carlo event generator and an event-by-event phase space analysis, we compute, for the first time, the production spectrum of \antihe and \antiT for dark matter annihilating or decaying to $b\bar{b}$ and ${W^+}{W^-}$ final states.  We then employ a semi-analytic model of interstellar and heliospheric propagation to calculate the \antihe flux as well as to provide tools to relate the anti-helium spectrum corresponding to an arbitrary antideuteron spectrum.  Finally, we discuss prospects for current and future experiments, including GAPS and AMS-02.
\end{abstract}

\maketitle

\section{Introduction}
\label{sec:intro}

Within the paradigm of Weakly Interacting Massive Particle (WIMP) dark matter, the pair-annihilation or decay of dark matter particles generically yields high-energy matter and antimatter cosmic rays. While the former are usually buried under large fluxes of cosmic rays of more ordinary astrophysical origin, antimatter is rare enough that a signal from dark matter might be distinguishable and detectable with the current generation of experiments. While astrophysical accelerators of high-energy positrons such as pulsars' magnetospheres are well-known, observations of cosmic anti-nuclei might provide a unique window into physics beyond the Standard Model and may provide a discovery route to unveil the nature of particle dark matter.

Measurements of the cosmic-ray antiproton spectrum by BESS~\cite{BESS,BESS2,BESS3} and PAMELA~\cite{PAMELA_ANTIPROTON} currently provide the best limits on cosmic-ray antiprotons \pbar in excess of the astrophysical background. On a short time-scale, AMS-02 will provide the most accurate cosmic-ray proton and antiproton spectrum to date, placing stringent limits on propagation parameters and excess signals.  One well motivated origin for such an excess is the annihilation or decay of WIMPs to hadronic final states -- generic to models coupling WIMPs to the weak gauge bosons or quarks (e.g. $W^+ W^-$ or $b\bar{b}$). While large astrophysical backgrounds often prohibit the clean disentanglement of exotic sources, a recent analysis projects that the 1-year AMS-02 data will produce robust constraints on WIMP annihilation to heavy quarks below the thermal-relic cross-section for dark matter masses $30 \leq m_{\chi} \leq 200 $ GeV~\cite{antiproton2}.

In addition to antiprotons, Ref.~\cite{donato2000} proposed new physics searches using heavier anti-nuclei such as antideuteron (\dbar), antihelium-3 (\antihe), or antitritium (\antiT) forming from hadronic neutralino annihilation products.  Although such production is of course highly correlated with the antiproton spectrum, the secondary astrophysical background decreases much more rapidly than the expected signal as the atomic number $A$ is increased~\cite{duperray2005}.  In particular, secondary antinuclei production from the spallation of high-energy cosmic rays -- i.e. the scattering of cosmic-ray protons off of cold interstellar hydrogen and helium -- quickly becomes kinematically suppressed for heavier nuclei for three reasons: 

(i) the constituent nucleons must lie in a small volume of phase space in order to form anti-nuclei, leading to a production suppression of roughly $10^{2A}-10^{3A}$.  While this is the case for both primary (e.g. dark matter) and secondary anti-nuclei, the secondary background is further suppressed by the rapid falloff of cosmic-ray protons at high energies.   The dominant spallation processes which generate \pbar, $\overline{\text{D}}$, and \antihe/\antiT have production thresholds of $7m_p$, $17m_p$, and $30m_p$ respectively while the proton flux above 10 GeV falls as $\phi_p \propto E^{-2.82}$~\cite{2011Sci}.  

(ii) because of the high production threshold, the spallation products are typically highly boosted,  carrying kinetic energies above 5 GeV/n (GeV per nucleon). For dark matter, the spectrum peaks instead below 1 GeV/n for annihilation channels where the hadronization frame is not boosted (e.g. $q\bar{q}$ or near threshold $W^+W^-$). 

(iii) finally, in contrast to $\bar{p}$, \dbar and \antihe easily fragment as they undergo inelastic collisions (due to their low binding energies).  This prevents efficient energy loss during interstellar transport which would otherwise redistribute the higher-energy background spectrum toward lower energies.

These three factors lead to precipitous decline in the secondary \dbar and \antihe backgrounds below $\sim 5$ GeV/n, enhancing the signal to background by several orders of magnitude for each increase in atomic number.  Proposed anti-nuclei searches exploit this point and are designed to observe below 1 GeV/n where the secondary/primary ratio for \antihe is $\lesssim 10^{-5}$.  This provides a truly zero-background channel for $A\geq 3$ at the expense of a significantly lower signal flux and it is precisely this feature which motivates dark matter searches using anti-nuclei.

Dark matter production of antideuterons and the observational prospects at AMS-02 and GAPS have been thoroughly investigated (see e.g. \cite{donato2000, profumo2005, cui2010,ibarraMC2013,kadastik,fornengo2013, ibarra2013}).  For an optimistic scenario of $\sim$100 GeV thermal WIMPs annihilating to $b \bar{b}$, the latter two state-of-the-art analyses predict $\mathcal{O}(0.1-10)$ \dbar signal events -- with backgrounds a factor $\mathcal{O}(10-50)$ smaller -- to be measured by a GAPS Long Duration Balloon flight (LDB+). It is thus naively expected that the extremely low \antihe flux will be difficult to observe.  While this is likely to be true for upcoming experiments, a future satellite based mission could potentially probe this zero-background channel.


The outline of this letter is as follows. In Section~\ref{sec:production} we discuss the coalescence model for the production of \antihe and calculate its formation rate relative to \dbar. In Section~\ref{sec:propagation} we employ a simple diffusion model in order to calculate the expected flux of \antihe at the solar position, and the penetration of \antihe into the heliosphere. In Section~\ref{sec:experiments} we discuss flux scaling relations, calculate the flux, and discuss the possibility for \antihe observation in both the current and upcoming AMS-02 and GAPS-LDB(+) experiments, as well as a future GAPS satellite mission. Finally in Section~\ref{sec:conclusions} we discuss the significance of our results to the current search for cosmic-ray anti-nucleons and conclude. 

\section{Dark Matter Production of Antihelium}
\label{sec:production}

We consider a fermionic Majorana dark matter candidate of mass $m_{\chi}$ annihilating into the colored or color-neutral final states $b\bar{b}$ and $W^+W^-$ through a generic, spin-0, $s$-channel resonance.  In the absence of an analytic description of atomic nuclei formation, we employ the coalescence model as a simple, single-parameter phenomenological approach to describe the formation of light elements from the distributions of protons and neutrons in high energy collisions~\cite{coal1,coal2}.  In the antideuteron case, the coalescence model assumes that nucleons with a relative invariant four-momenta $(k_n-k_p)^2=(\Delta \vec{k})^2 - (\Delta E)^2$ less than a coalescence momentum $p_0$, will bind together and form a nucleus. 

Early computations of the resulting antideuteron spectrum employed a fully factorized coalescence prescription in which the $\bar{p}$ and $\bar{n}$ momentum distributions were assumed to be uncorrelated and isotropic.  However, as demonstrated in Ref.~\cite{kadastik}, angular correlations introduced by jet structure play a crucial role in the formation of anti-nuclei, especially for heavy dark matter masses where the parton showers become increasingly focused.  Following more recent studies, we abandon the isotropic model and instead use the \texttt{PYTHIA 8.156}~\cite{pythia6,pythia8} event generator to simulate the parton shower and subsequent hadronization.  Using the phase space information from \texttt{PYTHIA}, we apply the coalescence prescription on an event-by-event basis, allowing for a full reconstruction of the nucleon distribution functions. In our Monte Carlo study, we also exclude contributions from baryons which are not spatially localized on the scale of the antidueteron's wave function (which spans $\sim2$ fm).  This is implemented by stabilizing particles with lifetime $\tau\gtrsim 2$ fm/c and, physically, stabilizes long-lived hadrons which decay weakly.  While this simultaneous localization in position and momentum space is of order the Heisenberg limit our results are insensitive to several order-of-magnitude variations of $\tau$, which is held fixed throughout this analysis.  This results from the significant gap between hadronic and weak decay timescales.  

The coalescence function has a single parameter, the coalescence momentum $p_0$, which must be fit to available collider data.  Following the approach of Refs. \cite{ibarra2013,fornengo2013,cui2010}, we use $e^+e^- \to$~\dbar measurements from ALEPH at the $Z^0$ resonance, finding $(5.9\pm 1.8 \pm .5) \times 10^{-6}$ antideuterons per hadronic $Z^0$-decay with \dbar momenta 0.62-1.03 GeV/c and polar angle $|\cos{\theta}|<0.95$ (\cite{aleph}).  Consistent with Refs. \cite{ibarra2013} and \cite{fornengo2013}, our Monte Carlo simulations reproduce this rate for a coalescence momentum $p_0^{A=2}=0.192\pm .030$ GeV/c.

For antihelium, the coalescence prescription is nearly identical.  When more than three particles are involved there are two obvious ways to define the coalescence mechanism.  One can either require that each of the relative momenta lie within a `minimum bounding momentum-sphere' of diameter $p_0^{A=3}$ (dubbed MBS here), or we can require that the relative invariant 4-momenta of each particle-pair is less than $p_0^{A=3}$ (dubbed particle-pairing or PP here).  If we consider a triangle with sides equal to the relative momenta of two particles, the two methods coincide for obtuse and right triangles.  For acute triangles, however, the value of $p_0^{A=3}$ required to form a nucleus can be up to 15\% larger than the PP case. The MBS prescription also avoids unnatural kinks in the required value of $p_0^{A=3}$ as the inclusive angle of this triangle is varied.  We therefore choose MBS which \emph{always} underestimates the yield with respect to the particle-pairing method for identical values of $p_0^{A=3}$. From a simple Monte-Carlo which assumes an isotropic distribution of nucleon momenta, we estimate that MBS produces only approximately 6\% fewer antihelium, although this difference becomes compounded exponentially for heavier elements.  Without an understanding of the strong dynamics of nuclear formation, it is not important to consider one method as `more accurate than another', but the difference should be kept in mind when comparing results between studies.

For nuclei of atomic number $A$, the coalescence model predicts a production rate $R(A) \propto p_0^{3(A-1)}$, making \antihe predictions particularly sensitive to nuclear physics uncertainties. The choice of coalescence momentum is known to have significant dependence on the details of the underlying scattering process and is measured to be larger for $A=3$ than $A=2$~\cite{heavyion}. While heavy-ion collisions provide the only available constraints on \antihe production, they do not resemble the dynamics of dark matter annihilation. In an attempt to bracket the effect of this uncertainty on the resulting \antihe spectrum, we derive values for the $A=3$ coalescence momentum, $p_0^{A=3}$, using two different methods.  In the first method, we choose to scale the antideuteron coalescence momentum, $p^{A=2}_0$, up to $p_0^{A=3}$ following the theoretically motivated scaling of Ref.~\cite{salati1997}, in which $p_0 \sim \sqrt{B}$ for total nuclear binding energy $B$:
\begin{eqnarray}
    p_0^{A=3}=\sqrt{B_{^3\overline{He}}/B_{\bar{D}}}\ p_0^{A=2}=0.357\pm 0.059 \text{ GeV/c.}
\end{eqnarray}
As a second method, we use heavy-ion results from the Berkeley Bevalac collider which fit \dbar, \antiT, and \antihe coalescence momenta for several collision species (C+C up to Ar+Pb) at incident energies from 0.4-2.1 GeV/n~\cite{heavyion}.  Averaging the measured $p^{A=3}_0/p^{A=2}_0$ (molecular targets excluded) we infer the relation
\begin{eqnarray}
p^{A=3}_0 = 1.28~p^{A=2}_0 = 0.246 \pm 0.038~\text{GeV/c}.
\end{eqnarray}
Without parton-level production rates, such as $pp \to$ \antihe at the LHC we need to rely on the outlined ad-hoc schemes, which yield the largest systematic uncertainty on the final flux.  In the remainder of this analysis, we use the binding energies to determine $p_0^{A=3}$.

Formation of antihelium-3 proceeds through two channels: directly through coalescence of $\bar{p}\bar{p}\bar{n}$, and through the formation and decay of tritium $(\bar{p}\bar{n}\bar{n})$.  As noted in Ref.~\cite{cosmic_antimatter}, the former channel is suppressed by the Coulomb repulsion of the antiprotons, while the tritium channel is not.  Although it is not clear what this suppression factor is, a conservative approach ignores the direct antihelium-3 channel completely. Tritium is stable on collider timescales, and therefore we can directly study the relative production rates.  Data from the Bevalac~\cite{heavyion} and CERN-SPS~\cite{heavyion2} heavy-ion collisions indicates that the ratio of tritium to antihelium-3 production rates $\epsilon = R_{\rm H3}/R_{\rm He3}$ varies between 0 and 1, perhaps as an increasing function the center of mass energy with efficiency near unity around $\mathcal{O}$(50 GeV).  For the rest of this analysis we choose $\epsilon=1$, but one may simply rescale dN/dE (or the final flux presented later) by a factor $(1+\epsilon)/2$ to regain full generality.  We note that this uncertainty is small compared to the weakly constrained coalescence momentum.

\begin{figure}[ht]
\begin{center}
\includegraphics[width=.5 \textwidth]{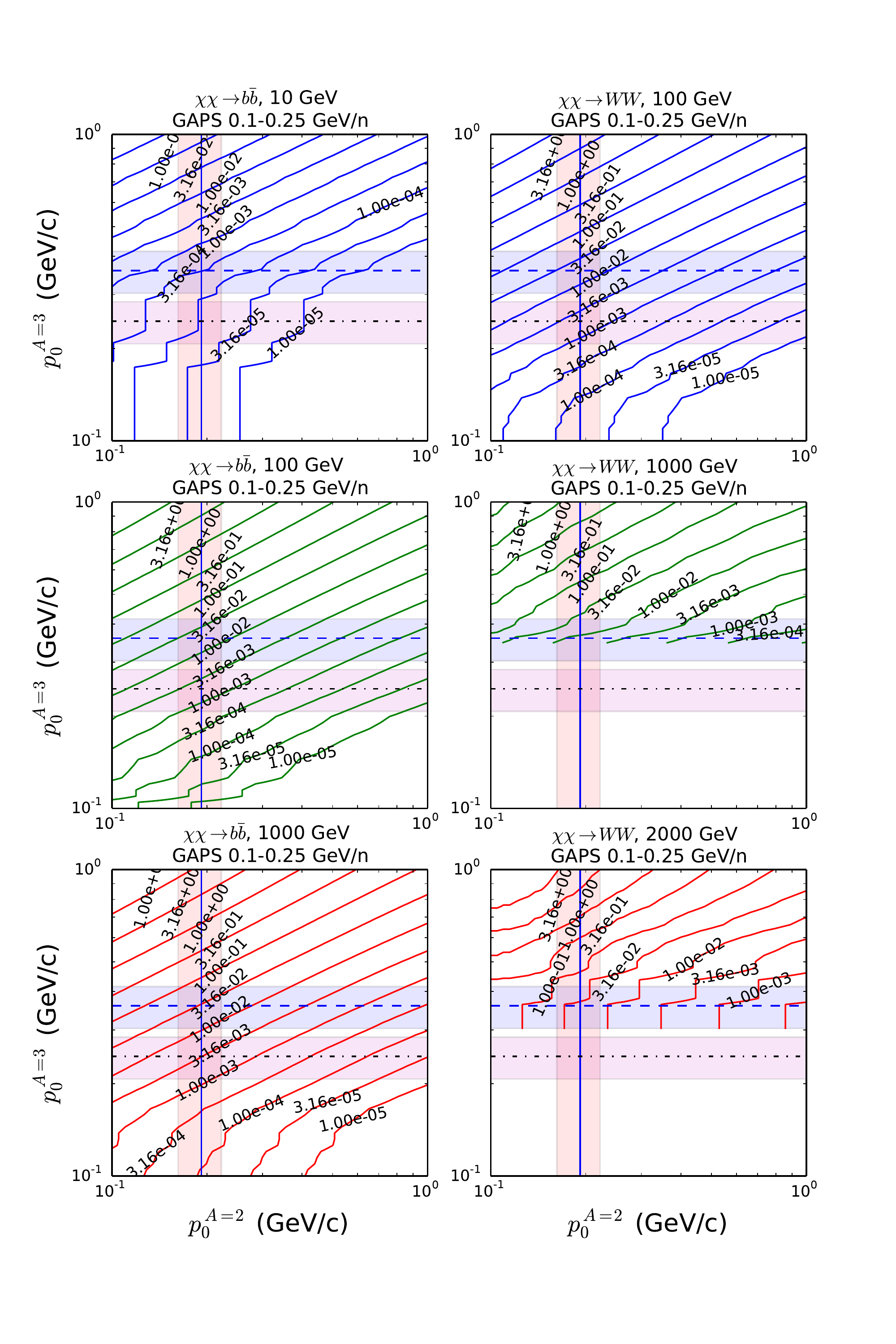}
\caption{ \footnotesize Ratios of the production of (\antihe + \antiT) to \dbar for Majorana dark matter annihilating to $b\bar{b}$ (left column) and $W^+W^-$ (right column) final states integrated over the energy bands for the proposed GAPS (LDB) instrument.  For each species, these bands were shifted for solar modulation according to a 500 MV Fisk potential.  The solid blue vertical lines show nominal values for $p_0^{A=2}$ with uncertainties (vertical shaded) while the horizontal lines show the $A=3$ coalescence momentum extrapolated using the nuclear binding energy (blue dashed) and heavy-ion data (black dot-dashed).  White regions with no contours contained no Monte-Carlo events.}
\label{fig:ratios}
\end{center}
\end{figure}

 In Figure~\ref{fig:ratios} we show ratios of the \antihe to \dbar injection spectra integrated over the energy band 0.1-0.25 GeV/n relevant for the upcoming GAPS Long Duration Balloon Flights (LDB and LDB+) as a function of the $A=2$ and $A=3$ coalescence momenta, for four different combinations of the dark matter pair-annihilation final state ($b\bar b$ in the left panels, $WW$ in the right panels) and mass (10, 1000 and 2000 GeV).  The GAPS energy bands are quoted for kinetic energies at the top of Earth's atmosphere, after the particle momenta have been shifted by propagation through the heliosphere.  Solar modulation will be discussed in detail in Section~\ref{sec:propagation}, but for concreteness, we integrate the \antihe and \dbar yields over bands shifted according to a Fisk potential of 500 MV in Figure~\ref{fig:ratios}.


The uncertainties on the coalescence momentum for $A=2$ are represented by the vertical shaded bands. For $A=3$ coalescence momenta, the two horizontal lines in each panel represent scaling with the binding energy (blue-dashed line) and heavy-ion collisions (black dot-dashed).  Regions with no visible contours produced no antihelium in the $2\times 10^{10}$ annihilation events simulated while the `wavy' lines are due to limited Monte Carlo statistics. We see that for most masses and final states that are potentially detectable (see discussion in Section~\ref{sec:experiments}) one should expect $10^{-3}-10^{-2}$ antihelium for each detected antideuteron. In the case of 10 GeV annihilation to b-quarks, the ratio is slightly lower as antihelium with a GAPS detectable kinetic energy requires a total energy of around 4.5 GeV.  However, this quickly increases toward the higher mass results as the dark matter mass is increased away from this threshold.  The effects induced by propagation of \dbar relative to \antihe are explored in the next section, but are sub-dominant compared with the nuclear physics uncertainties here.  In Sec.~\ref{sec:experiments} we  compute the actual flux and determine the detection prospects for future experiments.

\section{Astrophysical Production and Propagation of Antihelium}
\label{sec:propagation}

\subsection{The Dark Matter Source Term}

In order to create a Galactic model for dark matter annihilation throughout the galaxy which will allow us to map the \antihe injection spectrum to the flux at Earth, we must assume a dark matter halo model, a WIMP annihilation cross-section and a model of cosmic-ray transport. As a benchmark model, we choose a Navarro-Frenk-White (NFW) dark matter density profile, noting that Einasto and cored-isothermal profiles produce nearly identical results for the \dbar case \citep[see the discussion in][]{ibarra2013}:
\begin{eqnarray}
\rho_{\text{DM}}(r) = \rho_0 \left( \frac{r_s}{r} \right)^\alpha \frac{1}{(1+r/r_s)^{\alpha+1}}
\end{eqnarray}
with inner-slope $\alpha=1$, scale radius $r_s = 24.42$ kpc, and $\rho_0$ chosen to reproduce the dark matter density $\rho_{\odot} = 0.39$ GeV/cm$^3$ at the solar radius $r = 8.5$ kpc~\citep{2010JCAP...08..004C}.  For dark matter annihilation at a position $\vec{r}$, the source term for antihelium may then be written as
\begin{eqnarray}
Q_{\overline{\text{He}}}(T, \vec{r})=\frac{1}{2}\frac{\rho^2_{\text{DM}}(\vec{r})}{m_{\chi}^2} \langle \sigma v \rangle (1+\epsilon)\frac{dN_{\overline{\text{H3}}}}{dT},
\end{eqnarray}
where the $dN/dT$ term is the injection spectrum for tritium found in Section~\ref{sec:production}, $\epsilon$ is the ratio of the production rates of antihelium to tritium  (we take $\epsilon$=1 as discussed earlier), and  $\langle \sigma v \rangle$ is the thermal annihilation cross-section.  The source term must then be propagated from the site of annihilation to Earth.  This is typically broken down into two components: (i) interstellar propagation in which the cosmic-rays interact with turbulent Galactic magnetic fields, the interstellar hydrogen and helium, and Galactic winds, and (ii) propagation through the heliosphere, which can significantly deplete the low energy flux as the solar wind deflects charged particles.

\subsection{Propagation Models}
Interstellar propagation can be implemented via the well known stationary, cylindrically symmetric, two-zone diffusion model identical to the setup used for \dbar in Ibarra \& Wild \cite{ibarra2013} with the exception of obvious replacements including the \antihe cross-sections, charge, and atomic mass.  We assume a diffusion zone of radius 20 kpc and variable height $L$ with a thin, Galactic disk of half-height $h=100$ pc containing the interstellar medium.  The model is parametrized by an additional three components: an energy dependent diffusion constant $K(T)=K_0 ~ \beta ~\mathcal{R}^\delta$ with spectral index $\delta$, $\beta=v/c$ and rigidity $\mathcal{R}\equiv p(\text{GeV})/Z$ where Z is proton number, and $V_c$, which characterizes Galactic wind convection.  It is then possible to write the propagation in terms of the following transport equation:
\begin{eqnarray}
\begin{aligned}
\label{eq:transport}
\begin{split}
0=\frac{\partial n}{\partial t}=\nabla \cdot (K(T,\vec{r})~\nabla n) -\nabla\cdot (V_c ~\text{sign}(z)~\vec{k}~n) \\ -
2~h~\delta(z)~\Gamma_{\text{int}}~n + Q_{\overline{\text{He}}}(T,\vec{r}).
\end{split}
\end{aligned}
\end{eqnarray}

Here $n(T,\vec{r})$ is the antihelium number density and $\Gamma_{\mathrm{int}}$ is the interaction rate for antihelium within the ISM, described thoroughly in \S~\ref{subsec:cross-sections}.

The four parameters $L, K_0, \delta,$ and $V_c$ are then varied over the space consistent with the measured ratio of boron to carbon, with values producing the MIN/MED/MAX flux tabulated in Ref.~\cite{prop_model}.  The resulting uncertainty in the flux spans three orders of magnitude.  However, the \dbar and \antihe fluxes are tightly correlated to \pbar whose flux is well measured by PAMELA.  The propagation uncertainty on the \emph{maximal} \dbar (and \antihe) flux allowed by the measured $p$/\pbar ratio is then reduced to within a factor 4 of the MED model~\cite{ibarra2013}\footnote{We emphasize that propagation parameters are still fit using B/C and \textit{not} to the measured $p$/\pbar ratio which is only used to constrain the maximal propagation model.}. Upcoming antiproton results from AMS-02 will tighten this upper-limit and the large nuclear physics will certainly dominate in the case of antihelium.  In particular, the \antihe flux is sensitive to the sixth power of $p_0^{A=3}$, making updated collider production rates for \dbar and \antihe a crucial factor in \emph{any} estimate of an anti-nucleon flux.

The flux at the solar system is found by numerically integrating the dark matter annihilation rates over the dark matter halo and solving the transport equation analytically. For local dark matter density $\rho_\odot$, dark matter mass $m_\chi$, and thermal cross-section $\langle \sigma v \rangle$, the antihelium flux at the boundary of the solar system is given by 
\begin{align}
\label{eq:pnum}
\begin{split}
\Phi^{\mathrm{IS}}_{\overline{He}} (T)= \left(\frac{\rho_0}{0.39~\mathrm{GeV cm}^{-2}}\right)^2 \left(\frac{100~\mathrm{GeV}}{m_\chi}\right)^2\\
\times \left(\frac{\langle \sigma v\rangle}{3\times 10^{-26} \mathrm{cm}^3/\mathrm{s}}\right) \cdot P_{\mathrm{num}}(T) \cdot \frac{dN(T)}{dT},
\end{split}
\end{align}
where $P_{\mathrm{num}}(T)$ is the energy dependent numerical output of the propagation code and dN/dT is the \antihe injection spectrum from Sec.~\ref{sec:production}.  

\begin{figure}[ht]
\begin{center}
\includegraphics[width=.45 \textwidth]{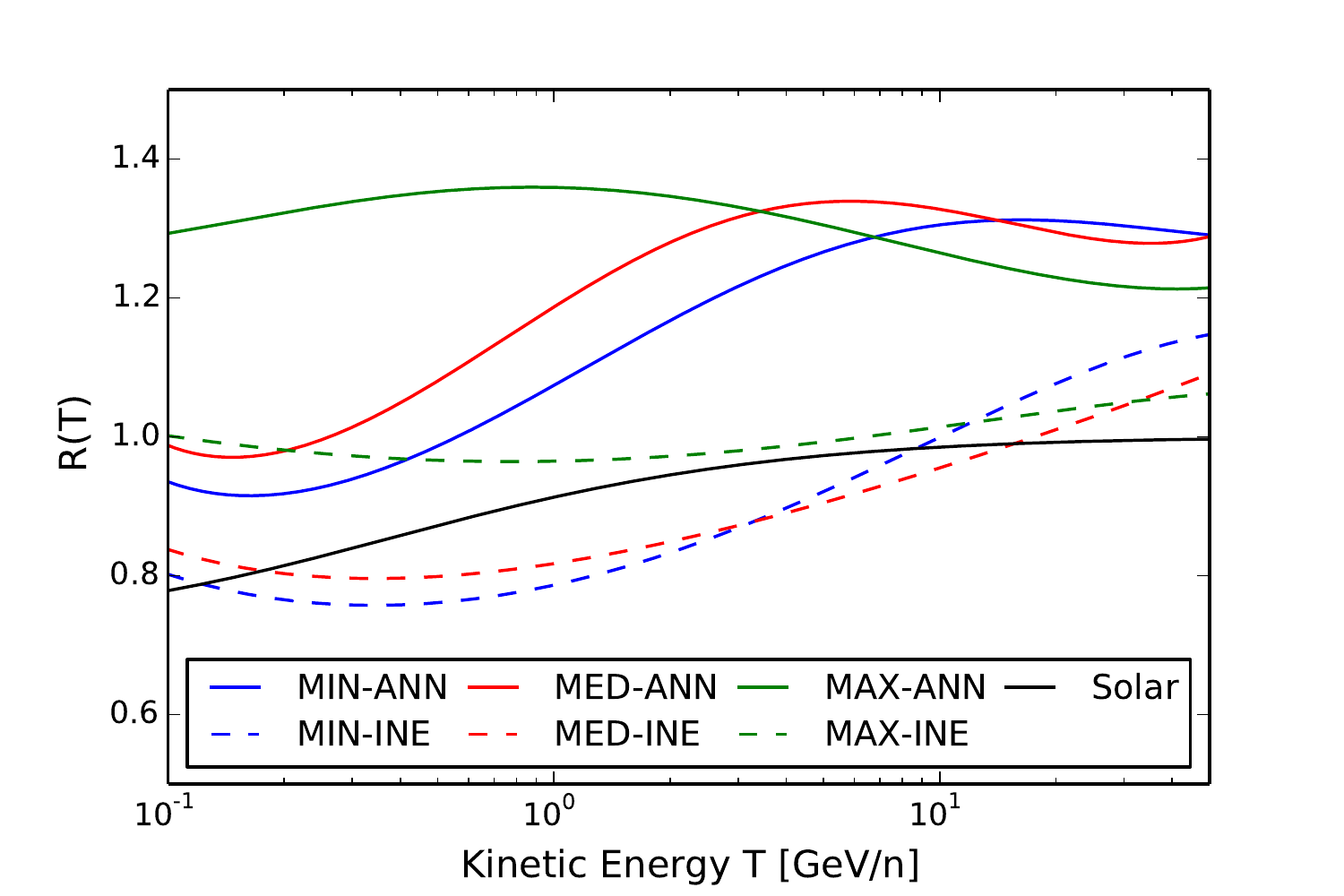}
\caption{ \footnotesize Propagation ratios $R(T)=P_{\mathrm{num}}^{\overline{\mathrm{He}}}/P_{\mathrm{num}}^{\overline{\mathrm{D}}}$ for ($P_{\mathrm{num}}$ in Eq.~(\ref{eq:pnum})) which show the enhancement or suppression of the antihelium flux with respect to antideuterons.  MIN/MED/MAX interstellar propagation models are shown in blue/red/green for two values of the propagation cross-section: Annihilation only (solid lines) and total-inelastic (dashed lines).  Also shown is the ratio of solar propagation functions in the Force Field Approximation (black solid).} 
\label{fig:propagation}
\end{center}
\end{figure}

In Figure~\ref{fig:propagation} we show the ratio $P_{\mathrm{num}}^{\mathrm{\overline{He}}}/P_{\mathrm{num}}^{\mathrm{\overline{D}}}$ for the MIN/MED/MAX propagation models and two values of the interaction rate, $\Gamma_{\mathrm{int}}$.  As we will discuss in \S~\ref{subsec:cross-sections}, uncertainty in the antihelium cross-section with interstellar gas can lead to a $\sim$25\% enhancement or suppression of the antihelium flux relative to that of antideuterons.  Of mild importance is the higher nuclear binding energy of \antihe compared to the very weakly bound \dbar case. While this can more efficiently deplete the higher energy population where the non-annihilating inelastic cross-section dominates, the low energies of interest here are not significantly enhanced by tertiary contributions which are ignored in our treatment.

In-fact, the two-zone diffusion model neglects all diffusion in momentum space, the most important of which may be a proper treatment of interstellar re-acceleration.  Several of these schemes, including diffusive re-acceleration, have been applied to the propagation of elements in more sophisticated numerical codes.  While these attempts have been successful in reproducing otherwise anomalous peaks in the secondary to primary ratios of heavy elements such as B/C, they encounter problems for light elements.  In particular, diffusive re-acceleration results in a spectral bump near 2 GeV/n for p and He which is not observed and the primary injection spectra must be artificially broken to compensate. This leads to an overestimate of the primary p and He flux by a factor $\sim$2~\cite{moskalenko2001_2}.  As we are concerned with light \& low energy nuclei, and no consensus on re-acceleration has been reached for this regime, we proceed without incorporating any re-acceleration mechanism.  This results in a primary spectrum within 20\% of measurements at low energies~\cite{moskalenko2001_2}.

The second phase of propagation is through the heliosphere and is computed using the Force Field Approximation of Gleeson \& Axford~\cite{forcefield}.  The flux at the top of the atmosphere is given by 
\begin{align}
\label{eq:solar}
\Phi^{\text{TOA}}_{\text{A,Z}} (T_{\text{TOA}}) = \left( \frac{2 m_A A~ T_{\text{TOA}} + A^2~ T_{\text{TOA}}^2}{2 m_A A~T_{\text{IS}} + A^2~ T_{\text{IS}}^2} \right)\Phi^{\text{IS}}_{A,Z}(T_{\text{IS}}),
\end{align}
where $m_A$ is the nucleus' mass, $T_{\text{IS}}$ is the kinetic energy per nucleon at the boundary of the solar system, $T_{\text{TOA}}$ is kinetic energy per nucleon at the top of Earth's atmosphere, and $T_{\text{IS}}=T_{\text{TOA}} + (e\phi_F |Z|/A)$.  The Fisk potential $\phi_F$ describes the strength of the solar modulation and varies over an 11 year cycle.  Here we take $\phi_F=500$~MV corresponding to the most optimistic detection scenario.  The ratio of the \antihe to \dbar case is shown in Figure~\ref{fig:propagation}. The lowered rigidity of \antihe causes a $\sim$50\% suppression at low energies relative to the \dbar modulation factor.  It has been shown that at GAPS energies, the Force Field Approximation is within a factor 2 of the minimum and maximum values computed in a full numerical treatment of heliospheric \dbar transport~\cite{fornengo2013}.  Much of the discrepancy between analytic and numerical models should disappear when taking the ratio of modulation between antihelium and antideuterons as the first order rigidity modifications are already captured by the Force-Field Model.

\subsection{Interaction Cross Sections}
\label{subsec:cross-sections}

In this subsection we discuss \antihe interaction rates with the ISM and compare them to the \dbar case.  $\Gamma_{\mathrm{int}}$ in Eq.~(\ref{eq:transport}) is given by:
\begin{eqnarray}
\label{eq:gamma_int}
\Gamma_{\text{int}}= (n_\text{H} + 4^{2/3} n_{\text{He}})~ v~ \sigma_{\overline{\text{He}},p}
\end{eqnarray}
where we have assumed the H and He gas cross-sections are related by a geometrical factor $4^{2/3}.$ 
For the Galactic Disk's interstellar hydrogen and helium densities we use $n_{\text{H}}= 1 ~\text{cm}^{-3}$ and $n_{\text{He}} = 0.07 n_{\text{H}} $. $v$ is the antihelium velocity through the ISM, and $\sigma_{\overline{\text{He}},p}$ is interaction cross-section of antihelium with protons.

Direct measurements of the antihelium-proton annihilation and inelastic cross-sections needed in Eq.~(\ref{eq:gamma_int}) are not available.  Instead, we use the parameterizations in from Moskalenko, Strong, \& Ormes~\cite{moskalenko2001} for the total inelastic, non-annihilating inelastic, and annihilation cross sections.  For an atomic nucleus ($A,|Z|$) impingent on a stationary proton with kinetic energy per nucleon $T$, these are given in mb by 
\begin{gather}
\begin{split}
\sigma^{\mathrm{tot}}_{\bar{p}A} = A^{2/3} [48.2+19 T^{-0.55} &+ (0.1-0.18T^{-1.2})Z \\
&+0.0012 T^{-1.5}Z^2]
\end{split}\\
\sigma^{\mathrm{ann}}_{\bar{p}A} = \sigma^{\mathrm{tot}}_{\bar{p}A}-\sigma^{\mathrm{non-ann}}_{\bar{p}A}\\ 
\sigma^{\mathrm{non-ann}}_{\bar{p}A} = \sigma^{\mathrm{inel}}_{pA}
\end{gather}
In the last equation, we assume that the non-annihilating inelastic cross-section for an antiproton-nucleus interaction is the same as the proton-nucleus interaction which can be well-approximated by
\begin{align}
\begin{split}
\sigma_{pA}^{\mathrm{inel}} &=45 A^{0.7} \left[1+0.016 \sin(5.3-2.63\ln A)\right] \\
&\times  \left\{
\begin{array}{ll}
1-0.62 e^{-T/0.2} \sin\left[\frac{10.9}{(10^3T)^{0.28}}\right], & T \leq 3;\\
1,                                                                  & T > 3;
\end{array} \right.\nonumber\\
\end{split}
\end{align}

\begin{figure}[!ht]
\begin{center}
\includegraphics[width=.50 \textwidth]{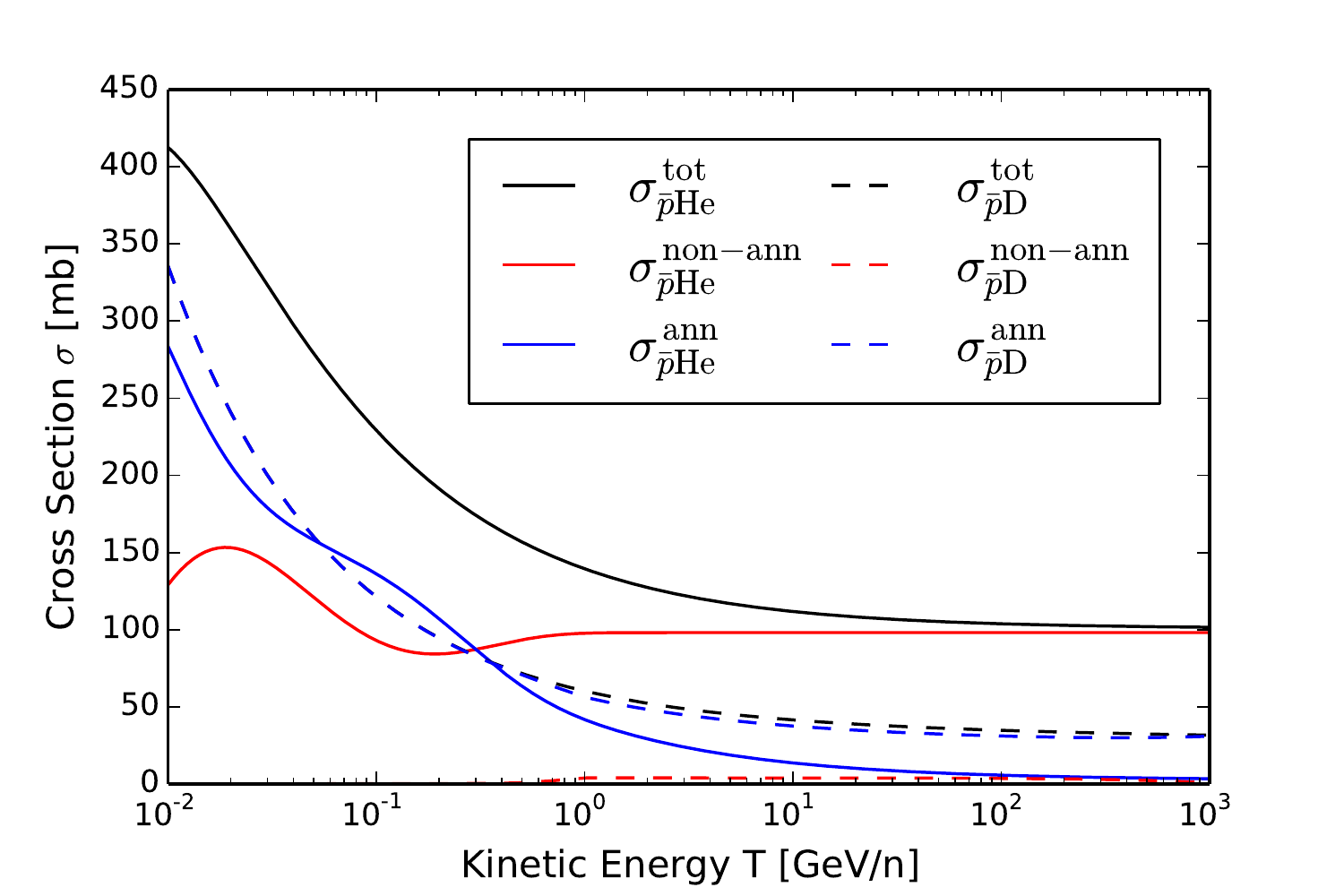}
\caption{ \footnotesize Proton--anti-nuclei inelastic scattering cross-sections as parametrized in Ref.~\cite{moskalenko2001}.  The non-annihilating inelastic cross-section for antideuterons is taken from Ref.~\cite{duperray2005}. Shown are the total inelastic (black), non-annihilating inelastic (red), and annihilation cross-sections for antihelium (solid) and antideuterons (dashed).}
\label{fig:cross-sections}
\end{center}
\end{figure}

In Figure~\ref{fig:cross-sections} we plot the three cross-sections for antihelium and antideuterons as a function of the kinetic energy per nucleon.  For the special case of \dbar, we take the parameterization from Tan \& Ng~\cite{tan1983} for total-inelastic cross-section, and an empirically determined non-annihilating inelastic cross-section which is very small due to the exceptionally low binding energy of \dbar \cite{duperray2005}.  Peaking at approximately 4 mb, this leads to a much higher probability of annihilation during inelastic scattering than the antihelium case.  We see that antihelium posesses an inelastic cross-section roughly 2 times larger than antideuterons at 1 GeV/n, while the opposite is true of the annihilation cross-sections.  In principle this implies a proportionally larger tertiary contribution for antihelium, where nuclear excitations remove kinetic energy during scattering.  In order to determine the relevance of this, one must also estimate the typical number of scatterings during propagation.  Assuming a cosmic-ray residence time $\tau_{\mathrm{res}} \approx 5\times10^6 ~\mathrm{yr}$ \cite{res_time}(which is only a weak function of rigidity, scaling at most as $\mathcal{R}^{-0.6}$)~\cite{res_time}, a mean hydrogen density $n_H=1$ cm$^{-2}$, and a typical interaction cross-section $\sigma\approx 100$ mb, the number of scatters can be found by comparing the residence path length $c~\tau_{\mathrm{res}}$ with the mean free path $\lambda$:

\begin{align}
N_{\mathrm{scatters}} = \frac{c \tau_{\mathrm{res}}}{\lambda}=c \tau_{\mathrm{res}}~n_H \sigma\approx 0.2.
\end{align}

With only a 20\% chance of scattering, and given the small amount of energy removed during the inelastic process, we ignore all tertiary contributions in our semi-analytic treatment of interstellar propagation.  

To bracket the impact of uncertainty in the  anti-nucleus -- proton cross-section, we use two methods: MethodANN and MethodINN which use the annihilation and total-inelastic cross-sections respectively in Eq.~(\ref{eq:gamma_int}). For \antihe, MethodINN leads to roughly a 40\% lower flux than MethodANN, while for \dbar, the results are nearly indistinguishable because of nearly identical total-inelastic and annihilation cross-sections. When examining the ratio of the resulting \antihe to \dbar flux, we see in Fig.~\ref{fig:propagation}, an enhancement (suppression) of order $25\%$  when using the annihilation (total-inelastic) cross-sections.


Now that the dark matter properties and propagation models have been fixed and the transport equation solved, we can translate the injection spectra calculated in Sec.~\ref{sec:production} into detectable fluxes at the top of the Earth's atmosphere.

\section{\antihe Flux and Detection Prospects for Current and Future Experiments}
\label{sec:experiments}

We have calculated injection spectra and propagation functions for \antihe, discussed the most important differences with respect to \dbar, and presented ratios for the conversion of \dbar spectrum into \antihe.  For concreteness, we reiterate the procedure here and show the most important scaling relations.  

With an antideuteron flux (or event rate) $\Phi_{\overline{D}}$ calculated within the coalescence framework described in Sec.~\ref{sec:production}, the antihelium flux is related through the following equation:
\begin{equation}
\begin{split}
\Phi_{\overline{He}} (T_{\mathrm{TOA}}) = R_{\textrm{IS}}(T_{\mathrm{IS}}) \cdot R_{\textrm{solar}}(T_{\mathrm{IS}})\left(\frac{p_0^{A=3}}{\overline{p}_{A=3} }\right)^6 \times \\
\times \left(\frac{{\overline{p}_{A=2}}}{p_0^{A=2}} \right)^3 \cdot R_{\text{PP}}(T_{\mathrm{IS}},m_\chi,f) \cdot \Phi_{\overline{D}} (T_{\mathrm{IS}}-e\phi_F/2),
\end{split}
\end{equation}

where $\overline{p}_{A=3}$~=~0.357~GeV/c  and $\overline{p}_{A=2}$~=~0.192~GeV/c. Here, $T_{\mathrm{IS}}=T_{\mathrm{TOA}}+(2/3)~e\phi_F$. $R_{\mathrm{PP}}$ is the particle production ratio, shown for GAPS energies from Fig.~\ref{fig:ratios} for the benchmark coalescence momenta. It is only a weak function of energy for the low energies relevant to these studies. $R_{\mathrm{IS}}(T_{\mathrm{IS}})$ and  $R_{\mathrm{solar}}(T_{\mathrm{IS}})$ are interstellar propagation ratios and the shifted solar ratios shown in Fig.~\ref{fig:propagation}.  This expression allows one to easily take more detailed analyses of \dbar spectra, rates or counts (as found in, for example, Refs.  \cite{ibarra2013, fornengo2013}) and scale them to the \antihe case, as well as incorporate new coalescence momentum measurements when they become available. 

We then compute the flux at the top of Earth's atmosphere for a  set of benchmark cases using the same dark matter models we considered in Sec.~\ref{sec:production} and the propagation setup described in Sec.~\ref{sec:propagation}.  In particular, we adopt $p_0^{A=2}=0.192, p_0^{A=3}=0.357$, MED propagation parameters, and use the slightly more optimistic ``MethodANN'' value for the antihelium interaction cross-section with the ISM.

\begin{figure}
\begin{center}
\includegraphics[width=.45 \textwidth]{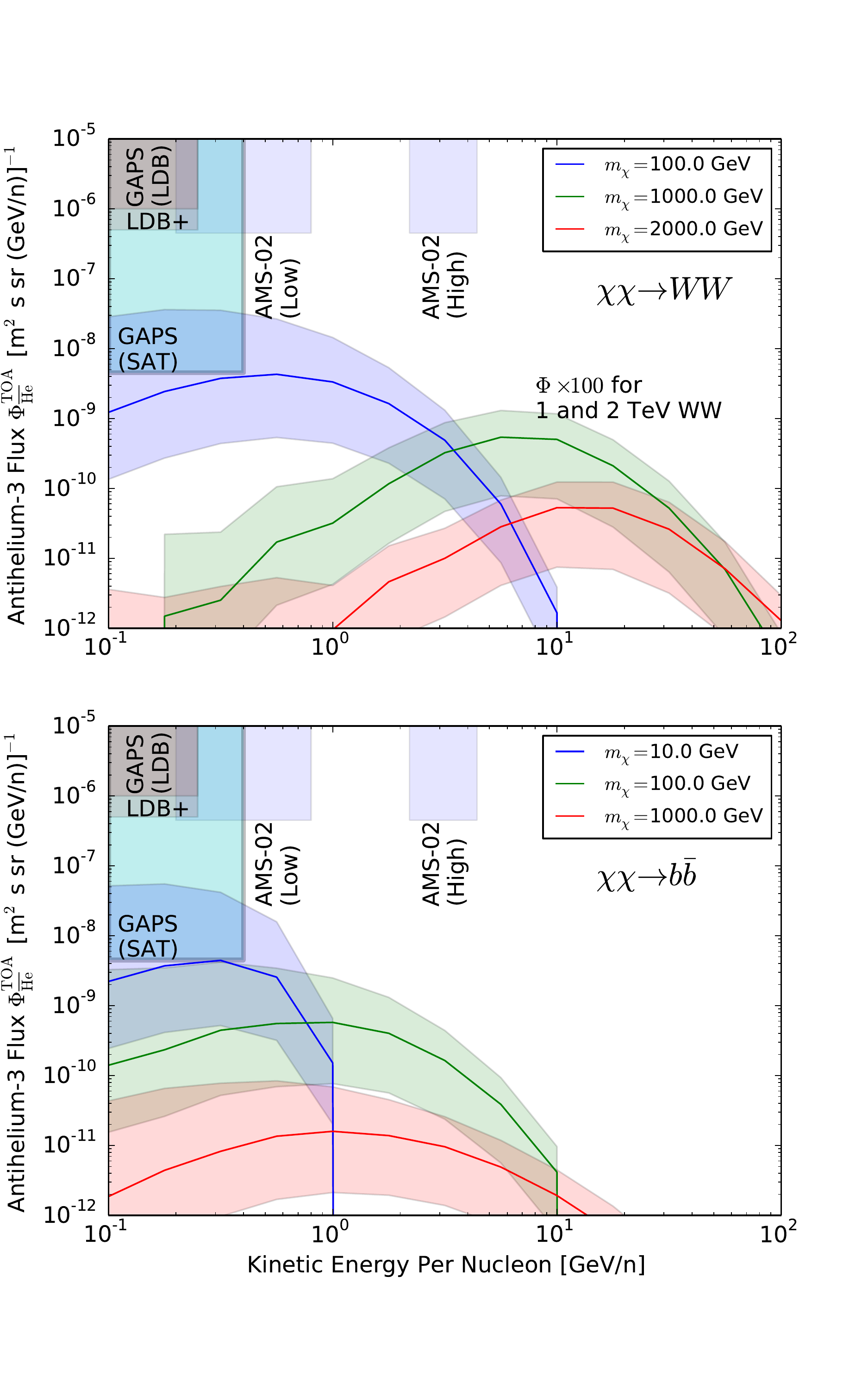}
\caption{ \footnotesize Flux of \antihe at the top of the atmosphere produced by dark matter annihilating to $W^+W^-$ (top) and  $b\bar{b}$(bottom) final states assuming an NFW dark matter density profile.  Flux is multiplied by 100 for $W^+W^-$ with $m_\chi=$1, 2 TeV. The shaded vertical bands represent the energy-bands and proposed sensitivities for various GAPS and AMS-02 observations.  Shaded uncertainty bands represent the MIN/MAX interstellar propagation models, although these are reduced to within a factor 4 of the central value after applying constraints from PAMELA measurements of the \pbar spectrum.  Nuclear physics uncertainties are not shown.}
\label{fig:flux}
\end{center}
\end{figure}

In Figure~\ref{fig:flux} we present the flux at the top of the Earth's atmosphere for dark matter annihilating to $W^+W^-$ and $b\bar{b}$ final states with a thermally-averaged pair annihilation cross section $\langle\sigma v\rangle=3\times10^{-26}\ {\rm cm}^3/{\rm s}$ as well as propagation uncertainties.  Also shown are the latest sensitivities for AMS-02, GAPS(LDB/LDB+)~\cite{snowmass2013} and a GAPS(SAT) mission as proposed in Ref.~\cite{GAPS_first}.  We note that the propagation uncertainties largely cancel after applying \pbar constraints from PAMELA while the uncertainty in the $A=3$ coalescence momentum leads to a flux uncertainty of 1-3 orders of magnitude (not-shown), independent of \pbar constraints.  The astrophysical \antihe background peaks with a flux of $10^{-12}$ [m$^2$ s sr GeV/n]$^{-1}$ at approximately 20 GeV/n~\cite{duperray2005}.  This is off-scale over all energies shown and rapidly declines at lower energies. By 1 GeV/n the flux has already dropped by another factor $10^2$. Over the low energies covered by GAPS it can be considered zero relative to the primaries.

For the case of decaying dark matter, the flux can be easily estimated from the annihilation case by modifying terms in Eq.~(\ref{eq:pnum}). First, the squared terms become linear as the reaction rate now traces the dark matter density $\rho_{\rm DM}$ rather than $\rho_{\rm DM}^2$. The numerical factor and thermal cross-section can then be replaced by finding an `equivalent lifetime', $\tau$, which provides an average flux equal to the annihilation case (for $\langle \sigma v \rangle=3 \times 10^{-26} {\rm cm}^3/{\rm s}$).  The term containing  $\langle \sigma v \rangle$ is then replaced by $(\tau_0/\tau)$. As benchmarks, for dark matter decaying to $b\bar{b}$ with mass $m_{\chi}^{\rm dec}=20$ GeV we find $\tau_0 \approx 7.5 \times 10^{26} s$, while for dark matter decaying to $W^+W^-$ with mass $m_{\chi}^{\rm dec}=200$ GeV, $\tau_0 \approx 7.5 \times 10^{27} s$. Here we note that $m_{\chi}^{\rm dec} = 2m_{\chi}$.

In the case of annihilation to heavy quarks, the very recent analysis of Ref.~\cite{antiproton2} has updated antiproton constraints on WIMP annihilation to heavy quarks.  Specifically, a thermal WIMP annihilating to heavy quarks is ruled out by current Fermi and PAMELA measurements up to approximately 30 GeV while AMS-02 should probe a thermal cross-section up to $\sim$200 GeV very soon. The antiproton flux is a very important indicator which is directly correlated to the production of heavier anti-nuclei. However, the coalescence momentum for \dbar and \antihe can float independently of such measurements and it is therefore not unreasonable that a \dbar excess could be observed in-spite of an expected exclusion from antiprotons.  For antihelium, an antiproton constraints are even less direct than the case of \dbar due to the unconstrained coalescence momentum.

It is clear that the current generation of experiments is very unlikely to be sensitive to primary antihelium from dark matter annihilation.  Future generation satellite born experiments using a GAPS(SAT) detector, as initially proposed in Ref.~\cite{GAPS_first}, could potentially be sensitive to WIMPs annihilating to $W^+W^-$ near threshold and $b\bar{b}$ at $\lesssim 10$ GeV. Unfortunately, higher masses quickly become undetectable, particularly in the $W^+W^-$ case.  If a convincing \dbar signal is observed at GAPS or AMS-02, follow-up \antihe observations may be needed to confidently rule out misidentified astrophysical secondaries.

There are two important technical instrumental differences in \antihe detection compared to \dbar which are not incorporated into our analysis.  GAPS works by measuring X-ray cascades emitted during the formation of exotic atoms from antimatter and the gas target. This technique requires the particle to stop completely inside the detector, and the large volume and weight required could be prohibitive for satellite based missions.  This also reduces the high-energy acceptance for heavier nuclei such as helium.  Finally, searches at even lower energies increase the importance of geomagnetic field effects and would require a satellite very close to the geomagnetic poles.

\section{Discussion and Conclusions}
\label{sec:conclusions}
Due to the low production rate of cosmic-ray anti-nuclei in interstellar proton-gas interactions, the observation of such particles remains an intriguing avenue for a positive signal from dark matter annihilation. We have, for the first time, modeled the production rates of $A=3$ cosmic-ray antinuclei by employing the {\tt PYTHIA} event generator to reconstruct the angular distribution of baryons on an event-by-event basis. Noting that the larger binding energy of \antihe compared to \dbar theoretically motivates a larger coalescence momentum for \antihe, we have shown that the expected \antihe flux at the solar position lies significantly above the ``four order of magnitude'' suppression of $A=3$ anti-nuclei compared to $A=2$ anti-nuclei, which is naively expected by the coalescence model. While it is still likely that \dbar would be discovered well before \antihe, this analysis shows that observations of \antihe are both technically feasible for future experiments, and may be essential to confirm that any \dbar observation does, in fact, correspond to the discovery of a dark matter particle.

Using the known instrumental configurations of current experiments, we have also shown that  \antihe is not detectable by AMS-02, or the current configuration of GAPS LDB+. However, the signal can possibly be detected by a future GAPS satellite mission. Moreover, an observation of \dbar during either of the earlier missions will greatly constrain the parameter space of astrophysical propagation models, allowing for a more accurate forecast of the instrumental qualities necessary in order to detect the \antihe signal with a future satellite mission. \newline

\begin{acknowledgments}
\noindent  SP is partly supported by the US Department of Energy under contract DE-FG02-04ER41268. The simulations for this research were carried out on the UCSC supercomputer Hyades, which is supported by National Science Foundation (award number AST-1229745) and UCSC. TL is supported by the National Aeronautics and Space Administration through Einstein Postdoctoral Fellowship Award Number PF3-140110. AI and SW were partially supported by the DFG cluster of excellence "Origin and Structure of the Universe," the TUM Graduate School and the Studienstiftung des Deutschen Volkes.
\end{acknowledgments}

\bibliography{antihelium}

\end{document}